\documentstyle[12pt]{article}
\begin{document}
\begin{center}
{\Large Nuclear recoil corrections to the $2p_{\frac{3}{2}}$
state energy of hydrogen-like and high $Z$ lithium-like atoms
  in all orders in $\alpha Z$}\\

\end{center}
\begin{center}
{A.N.Artemyev, V.M.Shabaev, and V.A.Yerokhin}
\\
\end{center}

\begin{center}
{\it Department of Physics, St.Petersburg State University,}\\
{\it Oulianovskaya 1, Petrodvorets, St.Petersburg 198904, Russia}
\end{center}

Nuclear recoil corrections ...
\\

\newpage
\begin{abstract}
The relativistic nuclear recoil corrections to
 the energy of the $2p_{\frac{3}{2}}$ state
 of hydrogen-like and the $(1s)^{2}2p_{\frac{3}{2}}$ state of high $Z$
lithium-like atoms in all orders in
$\alpha Z$  are calculated.
The calculations are carried out using the B-spline method for the
Dirac equation.
  For low $Z$ the results
of the calculation are in good agreement with the
$\alpha Z$ -expansion results.
 It is found that
the  total nuclear
recoil contribution to the energy of the $(1s)^{2}2p_{\frac{3}{2}}-
(1s)^{2}2s$ transition in lithium-like
uranium constitutes $-0.09\,eV$.

\end{abstract}
\newpage
\section{Introduction}
Till recently accurate QED calculations of the nuclear recoil
corrections to atomic energy levels were of interest mainly in
connection with high precision measurements of the Lamb shift
in hydrogen [1,2]. In this case the nuclear recoil corrections
may be calculated in lowest orders in $\alpha Z$ ($\alpha$ is
the fine structure constant, $Z$ is the nuclear charge).
 However recent achievements in experimental investigations of
 highly charged ions [3] require  calculations
 of the nuclear recoil corrections  right up $Z=92$.
 In the last case the parameter $\alpha Z$ can no longer be
considered small and, so, the calculations without expansion
in $\alpha Z$ are required.

In our previous paper [4] we calculated
 the nuclear recoil corrections in all orders in $\alpha Z$
for the $1s,\,2s,\, 2p_{\frac{1}{2}}$ states of hydrogen-like
atoms and the $(1s)^{2}2s,\;(1s)^{2}2p_{\frac{1}{2}}$ states
of high $Z$ lithium-like atoms. We found that the nuclear recoil
contribution to the energy of the $(1s)^{2}2p_{\frac{1}{2}}$ -
$(1s)^{2}2s$ transition in lithium-like uranium constitutes
$-0.07\, eV$ and, so, is comparable with the uncertainty of
the experimental value of the transition energy: $280.59(10)\,eV$
[3]. We found also the nuclear recoil contribution, additional
to Salpeter's one, to the Lamb shift ($n=2$) of hydrogen to be
$-1.32(6)\,kHz$. This result is in good agreement with recent
analytical calculations of the $\frac{m^{2}}{M}(\alpha Z)^{6}
\ln{(\alpha Z)}$ and   $\frac{m^{2}}{M}(\alpha Z)^{6}$
corrections [5-8], according to which the terms of order
$\frac{m^{2}}{M}(\alpha Z)^{6}
\ln{(\alpha Z)}$ cancel each other [5,6], the contribution of order
$\frac{m^{2}}{M}(\alpha Z)^{6}$ is $-0.77\,kHz$ for the
$2s$ state [7] and $0.58\,kHz$ for the $2p_{\frac{1}{2}}$
state [8].

In the present paper we extend the results of [4] to the case
of the $2p_{\frac{3}{2}}$ state of hydrogen-like and high $Z$
lithium-like atoms.

The relativistic units $\hbar=c=1$ are used in the paper.
\section{Hydrogen-like atoms}

The complete $\alpha Z$-dependence expressions for the nuclear
recoil corrections to the energy levels of hydrogen-like atoms
were first derived by one of the authors of the present paper
[9] (a part of the expressions was found earlier by Braun [10]).
These expressions consist of three contributions: the Coulomb
contribution, the one-transverse-photon contribution, and
the two-transverse-photon contribution. For a state $a$ the
Coulomb contribution is given by
\begin{eqnarray}
\Delta E_{c}&=&\Delta E_{c}^{(1)}+\Delta E_{c}^{(2)}\,,\\
\Delta E_{c}^{(1)}&=&\langle a|\frac{{\bf p}^{2}}{2M}|a\rangle\,,\\
\Delta E_{c}^{(2)}&=&\frac{2\pi i}{M}\int_{-\infty}^{\infty}d\omega\,
\delta_{+}^{2}(\omega)\langle a|[{\bf p},V_{c}]G(\omega+\varepsilon_{a})
[{\bf p},V_{c}]|a\rangle\,,
\end{eqnarray}
where $|a\rangle$ is the unperturbed state of the Dirac electron
in the Coulomb field of the nucleus,
$V_{c}=-\frac{\alpha Z}{r}$ is the Coulomb potential of the nucleus,
${\bf p}$ is the momentum operator, $\delta_{+}(\omega)=\frac{i}{2\pi}
(\omega+i0)^{-1}$, $G(\omega)=(\omega-H(1-i0))^{-1}$ is the relativistic
Coulomb Green function,
$ H=\mbox{\boldmath $\alpha$}{\bf p}+\beta m +V_{c}\,$.
The one-transverse-photon contribution is
\begin{eqnarray}
\Delta E_{tr(1)}&=&\Delta E_{tr(1)}^{(1)}+\Delta E_{tr(1)}^{(2)}\,,\\
\Delta E_{tr(1)}^{(1)}&=&-\frac{1}{2M}\langle a|\Bigl({\bf D}(0){\bf p}+
{\bf p}{\bf D}(0)\Bigr)|a\rangle\,,\\
\Delta E_{tr(1)}^{(2)}&=&-\frac{1}{M}\int_{-\infty}^{\infty}d\omega\,
\delta_{+}(\omega)\langle a|\Bigl([{\bf p},V_{c}]G(\omega+\varepsilon_{a})
{\bf D}(\omega)\nonumber\\
&&-{\bf D}(\omega)G(\omega+\varepsilon_{a})[{\bf p},V_{c}]
\Bigr)|a\rangle\,,
\end{eqnarray}
where
\begin{eqnarray}
D_{m}(\omega)=-4\pi\alpha Z\alpha_{l}D_{lm}(\omega)\,,
\end{eqnarray}
$\alpha_{l}\;(l=1,2,3)$ are the Dirac matrices, $ D_{lm}(\omega)$ is
the transverse part of the photon propagator in the Coulomb gauge.
In the coordinate representation it is
\begin{eqnarray}
D_{ik}(\omega,{\bf r})=-\frac{1}{4\pi}\Bigl\{\frac
{\exp{(i|\omega|r)}}{r}\delta_{ik}+\nabla_{i}\nabla_{k}
\frac{(\exp{(i|\omega|r)}
-1)}{\omega^{2}r}\Bigr\}\,.
\end{eqnarray}
The two-transverse-photon
contribution is
\begin{eqnarray}
\Delta E_{tr(2)}=\frac{i}{2\pi M}\int_{-\infty}^{\infty}d\omega\,
\langle a|{\bf D}(\omega)G(\omega +\varepsilon_{a}){\bf D}(\omega)|a\rangle\,.
\end{eqnarray}

The terms $\Delta E_{c}^{(1)}$ and $\Delta E_{tr(1)}^{(1)}$ are leading
at low $Z$. These terms can easily be calculated by using
the virial relations for the Dirac  equation [11-13]. Such a calculation
 gives [9]

\begin{eqnarray}
\Delta E^{(1)}&\equiv&\Delta E_{c}^{(1)}+\Delta E_{tr(1)}^{(1)}=
\frac{m^{2}-\varepsilon_{a}^{2}}{2M}\,.
\end{eqnarray}
The $\alpha Z$-expansion of this term is given in [4].

The terms $\Delta E_{c}^{(2)},\;\Delta E_{tr(1)}^{(2)},$ and
$\Delta E_{tr(2)}$
 are given in the form that allows one to use the
relativistic Coulomb Green function for their calculation.
(As was found by Pachucki and Grotch [7], this form is
 convenient for the $\alpha Z$-expansion
calculations as well.)
In [4] we transformed these equations to the form that is more
convenient for calculations using the finite basis set methods
[14] and calculated them for the $1s$, $2s$, and $2p_{\frac{1}{2}}$
states by using the B-spline method for the Dirac equation
 [15,16]. The calculation for the
 $2p_{\frac{3}{2}}$ state considered here was carried out in
the same way.

Table 1 shows the results of the numerical calculation
for the $2p_{\frac{3}{2}}$ state expressed in terms of
the function $P(\alpha Z)$ defined by
\begin{eqnarray}
\Delta E^{(2)} = \Delta E_{c}^{(2)}+\Delta E_{tr(1)}^{(2)}+
\Delta E_{tr(2)}=
\frac{m}{M}\frac{(\alpha Z)^{5}}{\pi n^{3}}P(\alpha Z)mc^{2}
\end{eqnarray}
The functions $P_{c}$, $P_{tr(1)}$, and $P_{tr(2)}$  correspond to the
contributions $\Delta E_{c}^{(2)}$,
 $\Delta E_{tr(1)}^{(2)}$, and $\Delta E_{tr(2)}$,
respectively. For comparison,
the Salpeter's
contribution [17] is
\begin{eqnarray}
P_{S}(\alpha Z)&=&\frac{8}{3}\,0.030017-\frac{7}{18}
= -0.30884\,.
\end{eqnarray}
 The uncertainties given in the table  correspond
only  to  errors
of the numerical calculation. In addition, there is an uncertainty due to
deviation from the point single particle model of the nucleus
 used here.

To make a more detailed comparison with the $\alpha Z$-expansion calculations
we represent  the function $P(\alpha Z)$
in the form
\begin{eqnarray}
P(\alpha Z)&=&a_{1}+a_{2}\alpha Z+a_{3}(\alpha Z)^{2}+a_{4}(\alpha Z)^{3}
\end{eqnarray}
The coefficients $a_{i}$ can be calculated from our
numerical results for the $P(\alpha Z)$-function.
 Such a calculation
 using the values of the $P(\alpha Z)$-function for
$Z=$1,2,3,5,10,15 gives
\begin{eqnarray}
a_{1}=-0.30883,\;\;\;\;\;\;a_{2}=1.040.
\end{eqnarray}
The coefficient $a_{1}$ is in good agreement with
 the Salpeter's result given by the equation (12).
The coefficient $a_{2}$ coincides, within errors of
the numerical procedure, with
the corresponding coefficient ($a_{2}=1.047$)
obtained by Golosov et.al. [8] ( the related coefficient from the
contribution (10) is equal to zero).

According to our numerical calculations the contribution
of the difference between $\Delta E^{(2)}$ and Salpeter's
correction for the $2p_{\frac{3}{2}}$ state of hydrogen
constitutes $0.42(2)\,kHz$. Analytical result for the
$\frac{m^{2}}{M}(\alpha Z)^{6}$ correction found in [8]
gives $0.423\,kHz$. ( In addition to this correction, in [8]
the correction of order $\frac{m^{2}}{M}\alpha^{2}(\alpha Z)^{4}$
is calculated to be $0.014\,kHz$.)

Let us consider the nuclear recoil corrections
for hydrogen-like uranium. According to the formula (10)
the first correction  is
\begin{eqnarray}
\Delta E_{2p_{\frac{3}{2}}}^{(1)}=0.0664\,eV\,.
\end{eqnarray}
The second correction defined by (11) is
\begin{eqnarray}
\Delta E_{2p_{\frac{3}{2}}}^{(2)}=0.0014\,eV\,.
\end{eqnarray}
In the next section we use these results to find the total nuclear
recoil contribution to the energy of the $2p_{\frac{3}{2}}-2s$ transition
in lithium-like uranium.
\section{High Z lithium-like atoms}
The complete $\alpha Z$-dependence expressions for the nuclear recoil
corrections to the energy levels of high $Z$ few-electron atoms
were derived in [18].
These corrections are the sum of
 one- and two-electron contributions. The one-electron corrections
are obtained by summing all the one-electron contributions
 over all the one-electron states that are
occupied.
The two-electron corrections for the case considered here are
\begin{eqnarray}
\Delta E_{c}^{(int)}&=&-\frac{1}{M}\sum_{\varepsilon_{n}=\varepsilon_{1s}}
\langle a|{\bf p}|n\rangle \langle n|{\bf p}|a\rangle\,,\\
\Delta E_{tr(1)}^{(int)}&=&\frac{1}{M}\sum_{\varepsilon_{n}=
\varepsilon_{1s}}
\Bigl\{\langle a|
{\bf p}|n\rangle\langle n|
{\bf D}(\varepsilon_{a}-\varepsilon_{n})|a\rangle\\
&&+\langle a|{\bf D}(\varepsilon_{a}-\varepsilon_{n})
|n\rangle\langle n|{\bf p}|a\rangle\Bigr\}\,,\nonumber\\
 \Delta E_{tr(2)}^{(int)}&=&-\frac{1}{M}\sum_{\varepsilon_{n}
=\varepsilon_{1s}}\langle a|{\bf D}(\varepsilon_{a}-\varepsilon_{n}
)|n\rangle \langle n|{\bf D}(\varepsilon_{a}-\varepsilon_{n}
)|a\rangle \,.
\end{eqnarray}

The table 2 shows the results of the calculation of the corrections
(17),(18), and (19)   for the
$(1s)^{2}2p_{\frac{3}{2}}$ state expressed in terms of the function
$Q(\alpha Z)$ defined by
\begin{eqnarray}
\Delta E^{int}\equiv \Delta E_{c}^{(int)}+\Delta E_{tr(1)}^{(int)}+
\Delta E_{tr(2)}^{(int)}= -\frac{2^{9}}{3^{8}}\frac{m^{2}}{M}(\alpha Z)^{2}Q
(\alpha Z)\,.
\end{eqnarray}
Here we have taken into account the known non-relativistic limit
of this correction [19]. Within the $\frac{m^{2}}{M}(\alpha Z)^{4}$
approximation the function  $Q(\alpha Z)$ that we denote by
 $Q_{L}(\alpha Z)$
  is [20]
\begin{eqnarray}
Q_{L}(\alpha Z)&=&1+(\alpha Z)^{2}\Bigl(-\frac{13}{48}+\frac{1}{2}
\ln{\frac{27}{32}}\Bigr)
\,.
\end{eqnarray}
For comparison, this function
 is  given in the table as well.
 The functions $Q_{c}(\alpha Z)$,
 $Q_{tr(1)}(\alpha Z)$, and  $Q_{tr(2)}(\alpha Z)$
 correspond to
the corrections $\Delta E_{c}^{(int)}$, $\Delta E_{tr(1)}^{(int)}$, and
$\Delta E_{tr(2)}^{(int)}$, respectively.
In leading orders in $\alpha Z$ they are
\begin{eqnarray}
Q_{c}(\alpha Z)&=&1+(\alpha Z)^{2}\Bigl(\frac{17}{48}+\frac{1}{2}\ln{\frac
{27}{32}}\Bigr)
\,,\\
Q_{tr(1)}(\alpha Z)&=&\frac{5}{8}(\alpha Z)^{2}\,,\\
Q_{tr(2)}(\alpha Z)&=&-\frac{25}{256}(\alpha Z)^{4}\,.
\end{eqnarray}

For low $Z$, in addition to the corrections considered here,
the  Coulomb electron-electron interaction corrections  to the
non-relativistic nuclear recoil contribution must be calculated
separately.
The main contribution from these corrections is of order
$\frac{1}{Z}(\alpha Z)^{2}\frac{m^{2}}{M}$.

Let us find the total nuclear recoil contribution to the energy of the
$(1s)^{2}2p_{\frac{3}{2}}-(1s)^{2}2s$
 transition in lithium-like uranium. According to our calculation  the
term $\Delta E^{int}$ contributes
 $-0.0345\,eV$. Adding to this value the one-electron
contribution defined by (15) and (16) and using the related values for
the $2s$ state from [4]
 we find
$$
\Delta E_{(1s)^{2}2p_{\frac{3}{2}}}-\Delta E_{(1s)^{2}2s}=-0.094\,eV.
$$
This value is significant enough
to be included in accurate QED calculations of
the transition energy.
\section*{Acnowledgments}
We wish to thank Dr. S.G.Karshenboim for stimulating
discussions. Useful discussions with Dr. T.Beier are
gratefully acknowledged.
 The research described in this
publication was made possible in part by Grant No. NWU300  from the
International Science Foundation and Grant No. 95-02-05571a from
the Russian Foundation for Fundamental Investigations.
A.N.A. thanks I.V.Konovalov for financial support.

\newpage
\begin{table}
\caption{The  results of the numerical calculation of the one-electron
 nuclear
recoil corrections to the $2p_{\frac{3}{2}}$ state energy expressed in
terms of the function $P(\alpha Z)$ defined by equation (11).}
\begin{tabular}{|c|l|l|l|l|}  \hline
$Z$&$P_{c}(\alpha Z)$&$P_{tr(1)}(\alpha Z)$&$P_{tr(2)}(\alpha Z)$&
$P(\alpha Z)$\\ \hline
1&  0.0000&-0.1385(2)&-0.1630(3)&-0.3013(4)\\ \hline
5& -0.0001&-0.1230&-0.1493(1)&-0.2724(1)\\ \hline
10&-0.0004&-0.1037&-0.1338&-0.2379\\ \hline
15&-0.0008&-0.0841&-0.1198&-0.2047\\ \hline
20&-0.0012&-0.0644&-0.1069&-0.1726\\ \hline
25&-0.0017&-0.0446&-0.0950&-0.1413\\ \hline
30&-0.0021&-0.0245&-0.0841&-0.1107\\ \hline
35&-0.0026&-0.0042&-0.0741&-0.0081\\ \hline
40&-0.0031& 0.0163&-0.0649&-0.0517\\ \hline
45&-0.0035& 0.0372&-0.0567&-0.0230\\ \hline
50&-0.0040& 0.0585&-0.0496& 0.0050\\ \hline
55&-0.0044& 0.0802&-0.0432& 0.0326\\ \hline
60&-0.0049& 0.1025&-0.0379& 0.0597\\ \hline
65&-0.0054& 0.1254&-0.0337& 0.0863\\ \hline
70&-0.0058& 0.1490&-0.0307& 0.1125\\ \hline
75&-0.0063& 0.1735&-0.0289& 0.1384\\ \hline
80&-0.0068& 0.1990&-0.0285& 0.1638\\ \hline
85&-0.0072& 0.2256&-0.0294& 0.1889\\ \hline
90&-0.0078& 0.2534&-0.0319& 0.2138\\ \hline
92&-0.0080& 0.2649&-0.0334& 0.2236\\ \hline
95&-0.0083& 0.2827&-0.0361& 0.2383\\ \hline
100&-0.0088& 0.3135&-0.0422& 0.2625\\ \hline
\end{tabular}
\end{table}
\begin{table}
\caption{The  results of the numerical calculation of the nuclear
recoil correction $\Delta E^{(int)}$ for the $(1s)^{2}2p_{\frac{3}{2}}$
 state  of lithium-like ions expressed in terms of the function
$Q(\alpha Z)$ defined by equation (20). $Q_{L}(\alpha Z)$ is the
leading contribution defined by equation (21).}
\begin{tabular}{|c|l|l|l|l|l|}  \hline
$Z$&$Q_{c}(\alpha Z)$&$Q_{tr(1)}(\alpha Z)$&$Q_{tr(2)}(\alpha Z)$&
$Q(\alpha Z)$&$Q_{L}(\alpha Z)$\\ \hline
5& 1.00036&-0.00083&0.00000&0.99953&0.99953\\ \hline
10&1.00143&-0.00333&0.00000&0.99810&0.99811\\ \hline
15&1.00323&-0.00748&0.00001&0.99573&0.99574\\ \hline
20&1.00574&-0.01330&0.00004&0.99239&0.99242\\ \hline
25&1.00896&-0.02077&0.00010&0.98809&0.98816\\ \hline
30&1.01291&-0.02989&0.00021&0.98281&0.98295\\ \hline
35&1.01757&-0.04065&0.00039&0.97653&0.97679\\ \hline
40&1.02294&-0.05303&0.00065&0.96926&0.96969\\ \hline
45&1.02902&-0.06704&0.00101&0.96097&0.96163\\ \hline
50&1.03579&-0.08264&0.00150&0.95165&0.95264\\ \hline
55&1.04323&-0.09981&0.00212&0.94129&0.94269\\ \hline
60&1.05132&-0.11854&0.00289&0.92988&0.93179\\ \hline
65&1.06001&-0.13877&0.00382&0.91742&0.91995\\ \hline
70&1.06925&-0.16047&0.00488&0.90390&0.90716\\ \hline
75&1.07894&-0.18356&0.00608&0.88930&0.89343\\ \hline
80&1.08897&-0.20796&0.00738&0.87362&0.87875\\ \hline
85&1.09914&-0.23356&0.00873&0.85686&0.86312\\ \hline
90&1.10921&-0.26019&0.01006&0.83896&0.84654\\ \hline
92&1.11313&-0.27109&0.01057&0.83148&0.83964\\ \hline
95&1.11880&-0.28765&0.01127&0.81988&0.82901\\ \hline
100&1.12736&-0.31561&0.01224&0.79951&0.81054\\ \hline
\end{tabular}
\end{table}
\end{document}